\documentclass{article}
\usepackage{amsfonts}
\usepackage{amsmath}

\input{tcilatex}

\begin{document}

\title{$\mathcal{PT}$-symmetry and Integrability 
\footnote{To appear in Acta Polytechnica, Proceeding of the 
Micro conference {\it Analytic and algebraic methods II},
Doppler Institute, Prague, April 2007.} }
\author{\hspace{-1.2cm} Andreas Fring \\
{\small {\ \hspace{-1.2cm} Centre for Mathematical Science, City University,
Northampton Square, London EC1V 0HB, UK} }\\
{\small {\ \hspace{-1.2cm} Department of Physics, University of
Stellenbosch, 7602 Matieland, South Africa}}}
\maketitle

\begin{abstract}
We briefly explain some simple arguments based on pseudo Hermiticity,
supersymmetry and $\mathcal{PT}$-symmetry which explain the reality of the
spectrum of some non-Hermitian Hamiltonians. Subsequently we employ $%
\mathcal{PT}$-symmetry as a guiding principle to construct deformations of
some integrable systems, the Calogero-Moser-Sutherland model and the
Korteweg deVries equation. Some properties of these models are discussed.
\end{abstract}

\section{Introduction}

Non-Hermitian Hamiltonians with complex eigenvalue spectrum have been
studied almost since the formulation of quantum mechanics, most prominently
as consistent descriptions of dissipative systems resulting for instance
from channel coupling \cite{FW}. It is also known for a very long time that
many interesting non-Hermitian Hamiltonians with \textit{real} eigenvalue
spectrum result naturally in various circumstances. For instance, it was
argued more than thirty years ago that the lattice versions of Reggeon field
theory \cite{Regge3} 
\begin{equation}
\mathcal{H}=\sum\nolimits_{\vec{\imath}}\left[ \Delta a_{\vec{\imath}%
}^{\dagger }a_{\vec{\imath}}+iga_{\vec{\imath}}^{\dagger }(a_{\vec{\imath}%
}+a_{\vec{\imath}}^{\dagger })a_{\vec{\imath}}+\tilde{g}\sum\nolimits_{\vec{%
\jmath}}(a_{\vec{\imath}+\vec{\jmath}}^{\dagger }-a_{\vec{\imath}}^{\dagger
})(a_{\vec{\imath}+\vec{\jmath}}-a_{\vec{\imath}})\right] ~~  \label{Regge}
\end{equation}%
with $a_{\vec{\imath}}^{\dagger },a_{\vec{\imath}}$ being standard creation
and annihilation operators and $\Delta ,g,\tilde{g}\in \mathbb{R}$, possess
a real eigenvalue spectrum\footnote{%
I am grateful to John Cardy for pointing this out.} \cite{Regge}. The
reduction of the Hamiltonian in (\ref{Regge}) to a single lattice site in
zero transverse dimension \cite{Regge2} is very reminiscent of the so-called
Swanson model \cite{Swanson}, which results by replacing the interaction
term with a simpler bilinear expression $ga^{\dagger }a^{\dagger }+\tilde{g}%
aa$. The latter model serves currently as a concrete popular solvable model
to exemplify various general features related to the study of non-Hermitian
Hamiltonians \cite{Swanson,HJ,JM,ACIso,MGH}. Affine Toda field theory with
complex coupling constant is a very prominent class of field theoretical
models, which are argued \cite{Holl,David} to be consistent despite their
Hamiltonians being non-Hermitian. Besides the study of such explicit models
related to non-Hermitian Hamiltonians, in particular their spectral
properties, the question of how to formulate the corresponding quantum
mechanical description consistently was first addressed in \cite{Urubu}. The
useful insight of how to implement $\mathcal{PT}$-symmetry into this
formulation has been obtained thereafter \cite{Bender:2002vv}.

The current large interest in the subject of non-Hermitian Hamiltonian
systems was initiated about nine years ago \cite{Bender:1998ke} by the
surprising numerical observation that even the class of simple non-Hermitian
Hamiltonians 
\begin{equation}
\mathcal{H}=p^{2}-g(iz)^{N},  \label{1}
\end{equation}%
defined on a suitable domain, possesses a real positive and discrete
eigenvalue spectrum for integers $N\geq 2$ with $g\in \mathbb{R}$. Supported
by the numerous new results and insights (for some recent reviews see \cite%
{specialCzech,special,CArev,Bendrev}), which have been obtained since, the
natural question arises of how to construct non-Hermitan Hamiltonians with
real eigenvalue spectra in a more systematical way.

The question I would like to address in this talk is how this may be
achieved, in particular by generalizing some integrable models.

\section{Real spectra of non-Hermitian Hamiltonians}

The activities in spectral theory usually focus on normal or self-adjoined
operators in some Hilbert space. With regard to the remarks made in the
introduction we shall first briefly review some arguments which may be used
to explain the reality of the spectra of non-Hermitian Hamiltonians and
thereafter employ them to construct new models, which depending on the
argument used are guaranteed, or at least are likely, to have a real
eigenvalue spectrum.

\subsection{Pseudo-Hermiticity}

Since a Hermitian operator, say $h=h^{\dagger }$, is guaranteed to have real
eigenvalues, i.e.~$h\phi =\varepsilon \phi $ with $\varepsilon \in \mathbb{R}
$, one may trivially construct isospectral Hamiltonians by means of a
similarity transformation $H=\eta ^{-1}h\eta $, such that $H\Phi
=\varepsilon \Phi $ with $\Phi =\eta ^{-1}\phi $. When $\eta $ is a
Hermitian operator this implies that the conjugation of $H$ is simply
achieved by $H^{\dagger }=\eta ^{2}H\eta ^{-2}$. Such type of Hamiltonians
are denoted as pseudo Hermitian Hamiltonians \cite%
{Urubu,Mostafazadeh:2002hb,Mostafazadeh:2001nr,Mostafazadeh:2002id,Mostafazadeh:2003gz}%
. One of the immediate virtues of the aforementioned relations is that $\eta
^{2}$ can be used consistently as a metric operator.

Given a Hermitian Hamiltonian it is of course trivial to construct several
isospectral non-Hermitian Hamiltonians in this manner simply by computing $%
\eta ^{-1}h\eta \rightarrow H$ for some positive $\eta $. However, the
interesting situations arise when given simple non-Hermitian Hamiltonians,
such as for instance (\ref{Regge}) and (\ref{1}), possible together with the
knowledge that they possess a positive real spectrum, and one tries to
construct their Hermitian counterparts by seeking convenient Hermitian
operators $\eta $, such that $\eta H\eta ^{-1}\rightarrow h=h^{\dagger }$.
Unfortunately, this is only feasible in an exact manner in some very rare
cases \cite{Znojil:1999qt,Swanson,HJ,JM,ACIso,MGH} and mostly one has to
rely on perturbation theory, see e.g. \cite%
{Bender:2004sa,Mostafazadeh:2004qh,CA,ACIso,Can}. More awkward is the fact
that when given exclusively the non-Hermitian Hamiltonian $H$, there might
be several Hermitian Hamiltonians counterparts and the metric is therefore
not even uniquely determined. One may select out a particular metric by
specifying for instance at least one more observable \cite{Urubu} or the
spectrum.

\subsection{Supersymmetry}

Another standard procedure, which produces isospectral Hamiltonians is to
employ Darboux transformations or equivalently a supersymmetric quantum
mechanical construction \cite{Witten,Cooper}. For this one considers
Hamiltonians $\mathcal{H}$, which can be decomposed into the form 
\begin{equation}
\mathcal{H}=H_{+}\oplus H_{-}=Q\tilde{Q}\oplus \tilde{Q}Q~.  \label{H}
\end{equation}

As indicated in (\ref{H}) one assumes that the two superpartner Hamiltonians 
$H_{\pm }$ factor into the two supercharges $Q$ and $\tilde{Q}$, which
intertwine the Hamiltonians $H_{\pm }$ as $QH_{-}=H_{+}Q$ and $\tilde{Q}%
H_{+}=H_{-}\tilde{Q}$. Evidently the two charges commute with the
Hamiltonian $\mathcal{H}$, i.e.~$[\mathcal{H},Q]=[\mathcal{H},\tilde{Q}]=0$,
and thus the $sl(1/1)$ algebra constitutes a symmetry of $\mathcal{H}$. As
pointed out by various authors \cite%
{Cannata:1998bp,Andrianov:1998vy,Tkachuk:2001wt,Dorey:2001hi,Mostafazadeh:2002sk,Sinha:2004ma,Znojil:2004ge}%
, one does not require the\ Hamiltonians $H_{\pm }$ to be Hermitian, such
that we allow $H_{\pm }^{\dagger }\neq H_{\pm }$. The only constraints,
which are natural to impose when one wishes to make contact with the
pseudo-Hermitian treatment in the previous section, are that the individual
factors of $H_{\pm }$ are conjugated as \cite{Mostafazadeh:2002sk} 
\begin{equation}
Q^{\dagger }=\eta _{-}^{2}\tilde{Q}\eta _{+}^{-2}\text{\qquad and\qquad }%
\tilde{Q}^{\dagger }=\eta _{+}^{2}Q\eta _{-}^{-2},  \label{QQ}
\end{equation}%
where the operators $\eta _{\pm }$ are Hermitian $\eta _{\pm }^{\dagger
}=\eta _{\pm }$. As an immediate consequence of (\ref{QQ}), both
Hamiltonians $H_{\pm }$ in (\ref{H}) become pseudo-Hermitian and possess
Hermitian counterparts $h_{\pm }^{\dagger }=h_{\pm }$\ 
\begin{equation}
H_{\pm }^{\dagger }=\eta _{\pm }^{2}H_{\pm }\eta _{\pm }^{-2}\qquad
\Leftrightarrow \qquad h_{\pm }=\eta _{\pm }H_{\pm }\eta _{\pm }^{-1}.
\end{equation}%
By construction all four Hamiltonians $h_{\pm }$, $H_{\pm }$ are therefore
isospectral 
\begin{equation}
H_{\pm }\Phi ^{\pm }=\varepsilon \Phi ^{\pm }\qquad \text{and\qquad }h_{\pm
}\phi ^{\pm }=\varepsilon \phi ^{\pm }
\end{equation}%
and their corresponding wavefunctions are intimately related 
\begin{equation}
\Phi ^{+}=Q\Phi ^{-}=\eta _{+}^{-1}\phi _{+}\quad \text{and\quad }\Phi ^{-}=%
\tilde{Q}\Phi ^{+}=\eta _{-}^{-1}\phi _{-}.  \label{Q}
\end{equation}%
One may now characterize four qualitatively different cases depending on\
the properties of the Hermitian operators $\eta _{\pm }$ in (\ref{Q}),
namely $i)$ for generic $\eta _{\pm }$ we have isospectral quartets, $~ii)$
for generic $\eta _{+}$ and $\eta _{-}=\mathbb{I}$ and $iii)$ for generic $%
\eta _{-}$ and $\eta _{+}=\mathbb{I}$ \ we find isospectral triplets and
finally $iv)$ for $\eta _{\pm }=\mathbb{I}$ we have isospectral doublets.
The interesting cases $ii)$ and $iii)$, which an contain Hermitian
Hamiltonian, have been considered in \cite{Andrianov:1998vy}.

Next one needs to specify the explicit representation for the supercharges
in terms of the superpotential $W(x)$. Setting the parameter $\hbar
^{2}/2m=1 $, the simplest choices are differential operators of first order 
\begin{equation}
Q=\frac{d}{dx}+W\text{\qquad and\qquad }\tilde{Q}=-\frac{d}{dx}+W
\end{equation}
such that the two superpartner Hamiltonians may be written as 
\begin{equation}
H_{\pm }=-\Delta +W^{2}\pm W^{\prime }=-\Delta +V_{\pm }.
\end{equation}
Alternative choices with higher order differential operators are discussed
for instance in \cite{Andrianov:1994aj}. Assuming further that $H_{-}$
possesses a discrete spectrum $H_{-}\Phi _{n}^{-}=\varepsilon _{n}\Phi
_{n}^{-}$, one may adjust the energy scale such that $H_{-}\Phi _{m}^{-}=0$
for some chosen $m$. In order to single out this groundstate wavefunction we
denote it as $\psi _{m}:=\Phi _{m}^{-}=c\exp [-\int W_{m}dx]$, $c\in \mathbb{%
C}$. Consequently the superpartner potentials may be expressed in terms of
the groundstate wavefunctions and acquire the forms 
\begin{equation}
W_{m}=-\frac{\psi _{m}^{\prime }}{\psi _{m}},~~\quad V_{-}^{m}=\frac{\psi
_{m}^{^{\prime \prime }}}{\psi _{m}},~~\quad V_{+}^{m}=2\left( \frac{\psi
_{m}^{\prime }}{\psi _{m}}\right) ^{2}-\frac{\psi _{m}^{^{\prime \prime }}}{%
\psi _{m}}.  \label{WV}
\end{equation}
Therefore the Hamiltonians 
\begin{equation}
H_{\pm }^{m}=-\Delta +V_{\pm }^{m}+E_{m}=-\Delta +W_{m}^{2}\pm W_{m}^{\prime
}+E_{m}  \label{Hm}
\end{equation}
are isospectral 
\begin{equation}
H_{\pm }^{m}\Phi _{n}^{\pm }=E_{n}\Phi _{n}^{\pm }\text{\quad for }n>m.
\end{equation}
In order to disentangle the Hermitian from the non-Hermitian case, we
separate the superpotential into its real and imaginary part $W_{m}=w_{m}+i%
\hat{w}_{m}$ with $w_{m}=w_{m}^{\dagger }$, $\hat{w}_{m}=\hat{w}%
_{m}^{\dagger }$ and likewise for the groundstate energy $E_{m}=\varepsilon
_{m}+i\hat{\varepsilon}_{m}$. With these notations we can re-write (\ref{Hm}%
) as 
\begin{equation}
H_{\pm }^{m}=-\Delta +w_{m}-\hat{w}_{m}\pm w_{m}^{\prime }+\varepsilon
_{m}+i(2w_{m}\hat{w}_{m}\pm \hat{w}_{m}^{\prime }+\hat{\varepsilon}_{m})
\end{equation}
Clearly we encounter the situation $ii)$ or $iii)$ when 
\begin{equation}
w_{m}=(\mp \hat{w}_{m}^{\prime }-\hat{\varepsilon}_{m})/2\hat{w}_{m}\qquad 
\text{or\qquad }\hat{w}_{m}=0,
\end{equation}
respectively.

When given a Hamiltonian, irrespective of being Hermitian or non-Hermitian,
and at least one wavefunction, the exploitation of supersymmetry is a very
constructive procedure to obtain isospectral Hamiltonians, which could also
be Hermitian or non-Hermitian.

\subsection{$\mathcal{PT}$-symmetry}

A further very simple and transparent way to explain the reality of the
spectrum of some non-Hermitian Hamiltonians results when we encounter
unbroken $\mathcal{PT}$-symmetry, which in the recent context was first
pointed out in \cite{Bender:2002vv}. It means that both the Hamiltonian 
\textit{and} the wavefunction remain invariant under a simultaneous parity
transformation $\mathcal{P}:x\rightarrow -x$ and time reversal $\mathcal{T}%
:t\rightarrow -t$, that is we require 
\begin{equation}
\left[ H,\mathcal{PT}\right] =0\quad \text{and\quad }\mathcal{PT}\Phi =\Phi ,
\label{PT}
\end{equation}%
where $\Phi $ is a square integrable eigenfunction on some domain of $H$. It
is crucial to note that the $\mathcal{PT}$-operator is an anti-linear
operator, i.e.~it acts as $\mathcal{PT}(\lambda \Phi +\mu \Psi )=\lambda
^{\ast }\mathcal{PT}\Phi +\mu ^{\ast }\mathcal{PT}\Psi $ with $\lambda ,\mu
\in \mathbb{C}$ and $\Phi ,\Psi $ being some eigenfunctions. An easy way to
convince oneself of this property is to consider the standard canonical
commutation relation $[x,p]=i$. Since $\mathcal{PT}$:$x\rightarrow
-x,p\rightarrow p$, we require $\mathcal{PT}$:$i\rightarrow -i$ to keep this
relation invariant. Utilizing now \textit{both} relations in (\ref{PT}) and
the anti-linear nature of the $\mathcal{PT}$-operator, a very simple
argument leads to the reality of the spectrum 
\begin{equation}
\varepsilon \Phi =H\Phi =H\mathcal{PT}\Phi =\mathcal{PT}H\Phi =\mathcal{PT}%
\varepsilon \Phi =\varepsilon ^{\ast }\mathcal{PT}\Phi =\varepsilon ^{\ast
}\Phi .  \label{arg}
\end{equation}%
Whereas the first relation in (\ref{PT}) is usually trivial to check, the
second is in general difficult to access as one rarely knows all the
wavefunctions. In case it does not hold one speaks of a broken $\mathcal{PT}$%
-symmetry and the eigenvalues come in complex conjugate pairs. All arguments
in this subsection were essentially already known to Wigner in 1960 \cite{EW}
relating to anti-linear operators in a completely generic form. Noting that
the $\mathcal{PT}$-operator is an example of such an operator these ideas
have been revitalized in a modified form and developed further in the recent
context of the study of non-Hermitian Hamiltonians \cite{Bender:2002vv}.

\section{$\mathcal{PT}$-symmetry as a guiding principle to construct new
models}

If we now wish to construct new models with real eigenvalue spectra, we may
in principle use any of the previous arguments. Clearly the exploitation of $%
\mathcal{PT}$-symmetry on the level of the Hamiltonian is the most direct
and transparent way, as one can just read of this property immediately.
Thereafter one can write down some new $\mathcal{PT}$-symmetric Hamiltonians
by means of simple deformations, i.e.~replacing for instance the potential $%
V(x)$ by $V(x)f(ix)$, $V(x)f(ixp)$, $V(x)+f(ix)$ or $V(x)+f(ixp)$, etc. with 
$f$ being some arbitrary function. Clearly the Hamiltonians in (\ref{Regge})
and (\ref{1}) are of this type. Of course these new models are not
guaranteed to have real spectra as the second property in (\ref{PT}) might
be spoiled. Nonetheless, they have a high chance to describe non-dissipative
physics and are potentially interesting.

\subsection{$\mathcal{PT}$-symmetric extensions for multi-particle systems}

Basu-Mallick and Kundu \cite{Basu-Mallick:2000af} were the first to write
down some non-Hermitian extensions for some integrable many-particle
systems, i.e.~the rational $A_{\ell }$-Calogero models \cite{Cal2} 
\begin{equation}
\mathcal{H}_{BK}=\frac{p^{2}}{2}+\frac{\omega ^{2}}{2}\sum%
\nolimits_{i}q_{i}^{2}+\frac{g^{2}}{2}\sum\nolimits_{i\neq k}\frac{1}{%
(q_{i}-q_{k})^{2}}+i\tilde{g}\sum\nolimits_{i\neq k}\frac{1}{(q_{i}-q_{k})}%
p_{i}\quad   \label{BK}
\end{equation}%
with $g,\tilde{g}\in \mathbb{R},q,p\in \mathbb{R}^{\ell +1}$. There are some
immediate questions one may pose \cite{AF} with regard to the properties of $%
\mathcal{H}_{BK}$: i) How can one formulate $\mathcal{H}_{BK}$ independently
of the representation for the roots? ii) Can one generalize $\mathcal{H}_{BK}
$ to other potentials apart from the rational one? iii) Can one generalize $%
\mathcal{H}_{BK}$ to other algebras or more precisely Coxeter groups? iv) Is
it possible to include more coupling constants? and in particular v) Are the
extensions still integrable? It turns out that the answer to all these
questions become all quite simple when one realises that (\ref{BK})
corresponds in fact to the standard Calogero model simply shifted in the
momenta. This means the similarity transformation $\eta $ is simply the
translation operator in $p$-space.

In order to see this and to answer the above questions we ignore the
confining term in (\ref{BK}) by taking $\omega =0$ and re-write the
Hamiltonian as 
\begin{equation}
\mathcal{H}_{\mu }=\frac{1}{2}p^{2}+\frac{1}{2}\sum\nolimits_{\alpha \in
\Delta }g_{\alpha }^{2}V(\alpha \cdot q)+i\mu \cdot p,  \label{HHH}
\end{equation}%
where $\Delta $ is now any root system invariant under Coxeter
transformations, $\mu =1/2\sum\nolimits_{\alpha \in \Delta }\tilde{g}%
_{\alpha }f(\alpha \cdot q)\alpha $, $f(x)=1/x$ and $V(x)=f^{2}(x)$. We have
also introduced coupling constants $g_{\alpha },\tilde{g}_{\alpha }$ for
each individual root. The Hamiltonians $\mathcal{H}_{\mu }$ are meaningful
for any representation of the roots and all Coxeter groups. For a specific
choice of the representation for the roots, namely $\alpha _{i}=\varepsilon
_{i}-\varepsilon _{i+1}$ for $1\leq i\leq \ell $ with $\varepsilon _{i}\cdot
\varepsilon _{j}=\delta _{ij}$ and the Coxeter group, i.e.~$A_{\ell }$, we
recover the expression in (\ref{HHH}). To establish the integrability of
these models it is crucial to note the following not obvious property 
\begin{equation}
\mu ^{2}=\alpha _{s}^{2}\tilde{g}_{s}^{2}\sum\nolimits_{\alpha \in \Delta
_{s}}V(\alpha \cdot q)+\alpha _{l}^{2}\tilde{g}_{l}^{2}\sum\nolimits_{\alpha
\in \Delta _{l}}V(\alpha \cdot q)  \label{mu}
\end{equation}%
where $\Delta _{s},\Delta _{l}$ denotes the short and long roots,
respectively. For the details of the proof of this identity we refer to \cite%
{AF}. As a consequence (\ref{mu}), we may re-express $\mathcal{H}_{\mu }$ in
form of the usual Calogero Hamiltonian with shifted momenta together with
some redefinitions of the coupling constants 
\begin{equation}
\mathcal{H}_{\mu }=\frac{1}{2}(p+i\mu )^{2}+\frac{1}{2}\sum\limits_{\alpha
\in \Delta }\hat{g}_{\alpha }^{2}V(\alpha \cdot q),~~~\hat{g}_{\alpha
}^{2}=\left\{ 
\begin{array}{c}
g_{s}^{2}+\alpha _{s}^{2}\tilde{g}_{s}^{2}\quad \text{for }\alpha \in \Delta
_{s} \\ 
g_{l}^{2}+\alpha _{l}^{2}\tilde{g}_{l}^{2}\quad \text{for }\alpha \in \Delta
_{l}%
\end{array}%
\right. .  \label{hhh}
\end{equation}%
Therefore, upon the redefinition of the coupling constant, we may obtain $%
\mathcal{H}_{\mu }$ by a similarity transformation as $\mathcal{H}_{\mu
}=\eta ^{-1}h_{\text{Cal}}\eta $ with $\eta =e^{-x\cdot \mu }$. The results
of section 2.1 apply therefore and one may construct for instance the
corresponding wavefunctions by $\Phi _{\mu }=\eta ^{-1}\phi _{\text{Cal}}$.
Similarly one can establish integrability with the help of a Lax pair with a
shifted momentum. One may verify that 
\begin{equation}
L=(p+i\mu )\cdot H+i\sum\limits_{\alpha \in \Delta }\hat{g}_{\alpha
}f(\alpha \cdot q)E_{\alpha }\quad \text{and}\quad M=m\cdot
H+i\sum\limits_{\alpha \in \Delta }\hat{g}_{\alpha }f^{\prime }(\alpha \cdot
q)E_{\alpha }  \label{LM}
\end{equation}%
fulfill the Lax equation $\dot{L}=\left[ L,M\right] $, upon the validity of
the classical equation of motion resulting from (\ref{HHH}), where the Lie
algebraic commutation relations 
\begin{equation*}
\left[ H_{i},H_{j}\right] =0,~~\left[ H_{i},E_{\alpha }\right] =\alpha
^{i}E_{\alpha },~~\left[ E_{\alpha },E_{-\alpha }\right] =\alpha \cdot H,~~%
\left[ E_{\alpha },E_{\beta }\right] =\varepsilon _{\alpha ,\beta }E_{\alpha
+\beta }.
\end{equation*}%
are taken to be in the Cartan-Weyl basis, i.e.~they are normalized as $\func{%
tr}(H_{i}H_{j})=\delta _{ij}$, $\func{tr}(E_{\alpha }E_{-\alpha })=1$. The
vector $m$ can be expressed in terms of the structure constant $\varepsilon
_{\alpha ,\beta }$ and the potential in the usual fashion. We note that the
Lax equation is $\mathcal{PT}$-symmetric as $\mathcal{PT}$:$L\rightarrow
L,M\rightarrow -M$. Naturally the conserved charges $I_{k}=tr(L^{k})/2$,
notably the Hamiltonian $I_{2}$, have the same property.

Having established the integrability of the Calogero models one may address
the question ii) and try to extend these considerations to other potentials.
Allowing now $f(x)=1/\sinh x$ and $f(x)=1/\func{sn}x$, we obtain the
hyperbolic and elliptic case with $V(x)=f^{2}(x)$. The integrability is
guaranteed by means of the same Lax pairs (\ref{LM}). However, when
expanding the square in (\ref{hhh}) the resulting Hamiltonian is not quite
of the form (\ref{HHH}) 
\begin{equation}
\mathcal{H}_{\mu }=\frac{1}{2}p^{2}+\frac{1}{2}\sum\limits_{\alpha \in
\Delta }\hat{g}_{\alpha }^{2}V(\alpha \cdot q)+i\mu \cdot p-\frac{1}{2}\mu
^{2},  \label{gen}
\end{equation}%
because the identity (\ref{mu}) does not hold for the other potentials. This
means the Hamiltonians in (\ref{gen}) constitute non-Hermitian \emph{%
integrable} extensions for Calogero-Moser-Sutherland (CMS)-models for all
crystallographic Coxeter groups, including, besides the rational, also
trigonometric, hyperbolic and elliptic potentials. Dropping the last term
would break the integrability for the non-rational potentials.

\subsection{$\mathcal{PT}$-symmetric deformations of the Korteweg deVries
equation}

An even more popular integrable model than the CMS-model is one having the
Korteweg-de Vries (KdV) equation \cite{KdV} as equation of motion 
\begin{equation}
u_{t}+uu_{x}+u_{xxx}=0.  \label{0KdV}
\end{equation}%
This equation is known to remain invariant under $x\rightarrow
-x,t\rightarrow -t,u\rightarrow u$, i.e.~it is $\mathcal{PT}$-symmetric. By
the same recipe outlined above we may then carry out the following
deformation $u_{x}\rightarrow -i(iu_{x})^{\varepsilon }$ with $\varepsilon
\in \mathbb{R}$, which was originally performed for the second term in \cite%
{BBCF} and for the third term in \cite{AFKdV}, leading to the equations 
\begin{equation}
u_{t}-iu(iu_{x})^{\varepsilon }+~u_{xxx}=0\ \ ~\ \ \ \ \text{\ ~~}%
\varepsilon \in \mathbb{R}  \label{BKdV}
\end{equation}%
and 
\begin{equation}
u_{t}+uu_{x}+i\varepsilon (\varepsilon -1)(iu_{x})^{\varepsilon
-2}~u_{xx}^{2}+\varepsilon (iu_{x})^{\varepsilon -1}u_{xxx}=0,  \label{AF}
\end{equation}%
respectively. For model in (\ref{BKdV}) one can establish the following
properties: the Galilean symmetry is broken, the model possess two conserved
quantities in terms of infinite sums and exhibits steady state solutions.
However, it is unclear how $\mathcal{PT}$-symmetric can be utilized further.
Instead (\ref{AF}), despite being more complicated, has some simpler
properties: it is Galilean invariant, possess three simple conserved
charges, exhibits steady state solutions, $\mathcal{PT}$-symmetry can be
utilized to explain the reality of the energy and it allows for a
Hamiltonian formulation with non-Hermitian Hamiltonian density 
\begin{equation}
\mathcal{H=}u^{3}-\frac{1}{1+\varepsilon }(iu_{x})^{\varepsilon +1}~\ \ ~\ \
\ \ \text{\ }\varepsilon \in \mathbb{R}\text{.}  \label{Haf}
\end{equation}%
Analogues of various different types of solutions of the KdV-equation have
been studied in \cite{BBCF,AFKdV}. No soliton solutions have been found and
it seems unlikely that the models are integrable.

\section{Conclusions}

We have demonstrated that $\mathcal{PT}$-symmetry serves as a very useful
guiding principle to construct new interesting models, some of which even
remain integrable. Being closely related to integrable models, these new
models have appealing features and deserve further investigation. Naturally
one may also reverse the setting and employ methods, which have been
developed in the context of integrable to address questions which arise in
the study of non-Hermitian Hamiltonians. For instance, one \cite{DDTrev} may
employ Bethe ansatz techniques to establish the reality of the spectrum for
Hamiltonians of the type (\ref{1}).

\noindent \textbf{Acknowledgments}: I would like to thank the members of the 
\v{R}e\v{z} Nuclear Physics Institute, especially Miloslav Znojil, for their
kind hospitality and for organizing this meeting. I also gratefully
acknowledge the kind hospitality granted by the members of the Department of
Physics of the University of Stellenbosch, in particular Hendrik Geyer. For
useful discussions I thank Carla Figueira de Morisson Faria and Paulo Gon%
\c{c}alves de Assis.


\begin{thebibliography}{10}

\bibitem{FW}
H.~Friedrich and D.~Wintgen,
\newblock Interfering resonances and bound states in the continuum,
\newblock Phys. Rev. {\bf A32}, 3231--3243 (1985).

\bibitem{Regge3}
J.~L. Cardy and R.~L. Sugar,
\newblock Reggeon field theory on a lattice,
\newblock Phys. Rev. {\bf D12}, 2514--2522 (1975).

\bibitem{Regge}
R.~C. Brower, M.~A. Furman, and K.~Subbarao,
\newblock Quantum spin model for Reggeon field theory,
\newblock Phys. Rev. {\bf D15}, 1756--1771 (1977).

\bibitem{Regge2}
J.~B. Bronzan, J.~A. Shapiro, and R.~L. Sugar,
\newblock Reggeon field theory in zero transverse dimensions,
\newblock Phys. Rev. {\bf D14}, 618--631 (1976).

\bibitem{Swanson}
M.~S. Swanson,
\newblock Transition elements for a non-Hermitian quadratic Hamiltonian,
\newblock J. Math. Phys. {\bf 45}, 585--601 (2004).

\bibitem{HJ}
H.~Jones,
\newblock On pseudo-Hermitian Hamiltonians and their Hermitian counterparts,
\newblock J. Phys. {\bf A38}, 1741--1746 (2005).

\bibitem{JM}
H.~Jones and J.~Mateo,
\newblock An Equivalent Hermitian Hamiltonian for the non-Hermitian $-x^4$
  Potential,
\newblock Phys. Rev. {\bf D73}, 085002 (2006).

\bibitem{ACIso}
C.~Figueira~de Morisson~Faria and A.~Fring,
\newblock Isospectral Hamiltonians from Moyal products,
\newblock Czech. J. Phys. {\bf 56}, 899--908 (2006).

\bibitem{MGH}
D.~P. Musumbu, H.~B. Geyer, and W.~D. Heiss,
\newblock Choice of a metric for the non-Hermitian oscillator,
\newblock J. Phys. {\bf A40}, F75--F80 (2007).

\bibitem{Holl}
T.~Hollowood,
\newblock Solitons in affine Toda field theory,
\newblock Nucl. Phys. {\bf B384}, 523--540 (1992).

\bibitem{David}
D.~I. Olive, N.~Turok, and J.~W.~R. Underwood,
\newblock Solitons and the energy momentum tensor for affine Toda theory,
\newblock Nucl. Phys. {\bf B401}, 663--697 (1993).

\bibitem{Urubu}
F.~G. Scholtz, H.~B. Geyer, and F.~Hahne,
\newblock Quasi-Hermitian Operators in Quantum Mechanics and the Variational
  Principle,
\newblock Ann. Phys. {\bf 213}, 74--101 (1992).

\bibitem{Bender:2002vv}
C.~M. Bender, D.~C. Brody, and H.~F. Jones,
\newblock Complex Extension of Quantum Mechanics,
\newblock Phys. Rev. Lett. {\bf 89}, 270401(4) (2002).

\bibitem{Bender:1998ke}
C.~M. Bender and S.~Boettcher,
\newblock Real Spectra in Non-Hermitian Hamiltonians Having PT Symmetry,
\newblock Phys. Rev. Lett. {\bf 80}, 5243--5246 (1998).

\bibitem{specialCzech}
M.~Znojil~(guest editors),
\newblock Special issue: Pseudo-Hermitian Hamiltonians in Quantum Physics,
\newblock Czech. J. Phys. {\bf 56}, 885--1064 (2006).

\bibitem{special}
H.~Geyer, D.~Heiss, and M.~Znojil~(guest editors),
\newblock Special issue dedicated to the physics of non-Hermitian operators
  (PHHQP IV) (University of Stellenbosch, South Africa, 23-25 November 2005),
\newblock J. Phys. {\bf A39}, 9965--10261 (2006).

\bibitem{CArev}
C.~Figueira~de Morisson~Faria and A.~Fring,
\newblock Non-Hermitian Hamiltonians with real eigenvalues coupled to electric
  fields: from the time-independent to the time dependent quantum mechanical
  formulation,
\newblock Laser Physics {\bf 17}, 424--437 (2007).

\bibitem{Bendrev}
C.~Bender,
\newblock Making Sense of Non-Hermitian Hamiltonians,
\newblock arXiv:hep-th/0703096.

\bibitem{Mostafazadeh:2002hb}
A.~Mostafazadeh,
\newblock Pseudo-Hermiticity versus PT symmetry. The necessary condition for
  the reality of the spectrum,
\newblock J. Math. Phys. {\bf 43}, 205--214 (2002).

\bibitem{Mostafazadeh:2001nr}
A.~Mostafazadeh,
\newblock Pseudo-Hermiticity versus PT-Symmetry II: A complete characterization
  of non-Hermitian Hamiltonians with a real spectrum,
\newblock J. Math. Phys. {\bf 43}, 2814--2816 (2002).

\bibitem{Mostafazadeh:2002id}
A.~Mostafazadeh,
\newblock Pseudo-Hermiticity versus PT-Symmetry III: Equivalence of
  pseudo-Hermiticity and the presence of anti-linear symmetries,
\newblock J. Math. Phys. {\bf 43}, 3944--3951 (2002).

\bibitem{Mostafazadeh:2003gz}
A.~Mostafazadeh,
\newblock Exact PT-Symmetry Is Equivalent to Hermiticity,
\newblock J. Phys. {\bf A36}, 7081--7092 (2003).

\bibitem{Znojil:1999qt}
M.~Znojil,
\newblock PT -symmetric harmonic oscillators,
\newblock Phys. Lett. {\bf A259}, 220--223 (1999).

\bibitem{Bender:2004sa}
C.~M. Bender, D.~C. Brody, and H.~F. Jones,
\newblock Extension of PT-symmetric quantum mechanics to quantum field theory
  with cubic interaction,
\newblock Phys. Rev. {\bf D70}, 025001(19) (2004).

\bibitem{Mostafazadeh:2004qh}
A.~Mostafazadeh,
\newblock PT-Symmetric Cubic Anharmonic Oscillator as a Physical Model,
\newblock J. Phys. {\bf A38}, 6557--6570 (2005).

\bibitem{CA}
C.~Figueira~de Morisson~Faria and A.~Fring,
\newblock Time evolution of non-Hermitian Hamiltonian systems,
\newblock J. Phys. {\bf A39}, 9269--9289 (2006).

\bibitem{Can}
E.~Caliceti, F.~Cannata, and S.~Graffi,
\newblock Perturbation theory of $\cal{PT}$-symmetric Hamiltonians,
\newblock J. Phys. {\bf A39}, 10019--10027 (2006).

\bibitem{Witten}
E.~Witten,
\newblock Constraints on supersymmetry breaking,
\newblock Nucl. Phys. {\bf B202}, 253--316 (1982).

\bibitem{Cooper}
F.~Cooper, A.~Khare, and U.~Sukhatme,
\newblock Supersymmetry and quantum mechanics,
\newblock Phys. Rept. {\bf 251}, 267--385 (1995).

\bibitem{Cannata:1998bp}
F.~Cannata, G.~Junker, and J.~Trost,
\newblock Schr{\"o}dinger operators with complex potential but real spectrum,
\newblock Phys. Lett. {\bf A246}, 219--226 (1998).

\bibitem{Andrianov:1998vy}
A.~A. Andrianov, F.~Cannata, J.~P. Dedonder, and M.~V. Ioffe,
\newblock SUSY Quantum Mechanics with Complex Superpotentials and Real Energy
  Spectra,
\newblock Int. J. Mod. Phys. {\bf A14}, 2675--2688 (1999).

\bibitem{Tkachuk:2001wt}
V.~M. Tkachuk and T.~V. Fityo,
\newblock Factorization and superpotential of the PT symmetric Hamiltonian,
\newblock J. Phys. {\bf A34}, 8673--8677 (2001).

\bibitem{Dorey:2001hi}
P.~Dorey, C.~Dunning, and R.~Tateo,
\newblock Supersymmetry and the spontaneous breakdown of PT symmetry,
\newblock J. Phys. {\bf A34}, L391--L400 (2001).

\bibitem{Mostafazadeh:2002sk}
A.~Mostafazadeh,
\newblock Pseudo-Supersymmetric Quantum Mechanics and Isospectral Pseudo-Hermi
  tian Hamiltonians,
\newblock Nucl. Phys. {\bf B640}, 419--434 (2002).

\bibitem{Sinha:2004ma}
A.~Sinha and P.~Roy,
\newblock Generation of exactly solvable non-Hermitian potentials with real
  energies,
\newblock Czech. J. Phys. {\bf 54}, 129--138 (2004).

\bibitem{Znojil:2004ge}
M.~Znojil,
\newblock PT-symmetric regularizations in supersymmetric quantum mechanics,
\newblock J. Phys. {\bf A37}, 10209--10222 (2004).

\bibitem{Andrianov:1994aj}
A.~A. Andrianov, F.~Cannata, J.~P. Dedonder, and M.~V. Ioffe,
\newblock Second order derivative supersymmetry, Q deformations and scattering
  problem,
\newblock Int. J. Mod. Phys. {\bf A10}, 2683--2702 (1995).

\bibitem{EW}
E.~Wigner,
\newblock Normal form of antiunitary operators,
\newblock J. Math. Phys. {\bf 1}, 409--413 (1960).

\bibitem{Basu-Mallick:2000af}
B.~Basu-Mallick and A.~Kundu,
\newblock Exact solution of Calogero model with competing long-range
  interactions,
\newblock Phys. Rev. {\bf B62}, 9927--9930 (2000).

\bibitem{Cal2}
F.~Calogero,
\newblock Solution of a three-body problem in one-dimension,
\newblock J. Math. Phys. {\bf 10}, 2191--2196 (1969).

\bibitem{AF}
A.~Fring,
\newblock A note on the integrability of non-Hermitian extensions of
  Calogero-Moser-Sutherland models,
\newblock Mod. Phys. Lett. {\bf 21}, 691--699 (2006).

\bibitem{KdV}
D.~J. Korteweg and deVries G.,
\newblock On the change of form of long waves advancing in a rectangular canal,
  and on a new type of long stationary waves,
\newblock Phil. Mag. {\bf 39}, 422--443 (1895).

\bibitem{BBCF}
C.~M. Bender, D.~C. Brody, J.~Chen, and E.~Furlan,
\newblock PT-Symmetric Extension of the Korteweg-de Vries Equation,
\newblock J. Phys. {\bf A40}, F153--F160 (2007).

\bibitem{AFKdV}
A.~Fring,
\newblock PT-Symmetric deformations of the Korteweg-de Vries equation,
\newblock J. Phys. {\bf A40}, 4215--4224 (2007).

\bibitem{DDTrev}
P.~Dorey, C.~Dunning, and R.~Tateo,
\newblock The ODE/IM Correspondence,
\newblock arXiv:hep-th/0703066 .

\end{thebibliography}

\small{

}

\end{document}